The One Dimensional *Approachissimo* Quantum Harmonic Oscillator: The Hilbert-Polya Hamiltonian for the Primes and the Zeros of the Riemann Function.

N. García
Laboratorio de Física de Sistemas Pequeños y Nanotecnología
Consejo Superior de Investigaciones Científicas
Serrano 144, Madrid 28006, Spain.
*(nicolas.garcia@fsp.csic.es)*

I have made an ample study of one dimensional quantum oscillators, ranging from logarithmic to exponential potentials. I have found that the eigenvalues of the hamiltonian of the oscillator with the limiting (*approachissimo*) harmonic potential $(V_R(x) \approx x/\log(x) \, J(x) \sim p(x)^2)$ maps the zeros of the Riemann function height up in the Riemann line. This is the potential created by the field of *J(x)* that is the Riemann generator of the prime number counting function, *p(x)*, that in turn can be defined by an integral transformation of the Riemann zeta function. This plays the role of the spring strength of the quantum limiting harmonic oscillator. *The number theory meaning of this result is that the roots height up of the zeta function are the eigenvalues of a Hamiltonian whose potential is the number of primes squared up to a given x. Therefore this may prove the never published Hilbert-Polya conjecture.* The conjecture is true but does not imply the truth of the Riemann hypothesis. We can have complex conjugated zeros off the Riemman line and map them with another hermitic operator and a general expression is given for that. The zeros off the line affect the fluctuation of the eigenvalues but not their mean values.

I. INTRODUCTION.

The proof of the Riemann hypothesis would be a remarkable result[1]. Solving this problem will close many existing hypothesis and conjectures. To illustrate the problem read, for example, the recent popular books [2, 3] among others. There is a conjecture by Hilbert and Polya [4] that has never been published, that says: "the zeros of the Riemann zeta function may correspond to the eigenvalues of a physical hermitian operator". This may or may not imply the Riemann hypothesis that states: "All the zeros of the Riemann zeta function, *?(z)*, fall in the line $1/2 + it$, where *i* is the imaginary unity". The complex function reads among other expressions as:

$$z(z) = \sum_n n^{-z} = \prod_p (1 - p^{-z})^{-1} \quad , \quad (\text{Re } z > 1) \qquad (1)$$

where *n* and *p* are the integer and prime numbers.

In addition, it was proven [1], that for a given value *T* up in the Riemann line the average number of Riemann zeros goes as,

<(Number of Riemann Zeros up to large T )> = <$N_R(T)$> = $T/2p(log(T/2p) – 1)$    (2)

Large scale numerical simulations by Odlyzco [5] obtained the first up $10^{11}$ zeros and this permits calculating statistical moments and calculated the variance of the zeros. This variance seems to fit that of a "Gaussian unitary ensemble" (GUE), the ensemble owing to a hermitian operator. Berry [6, 7] took these statistics and argued that this could be explained by a classical limit of a hamiltonian whose trajectories are chaotic and do not possess time-reversal symmetry. In fact Berry simulated it with random matrices for billiard confinements. This brings us to the idea that the zeros of the zeta function are the eigenvalues of such a classic chaotic system. And the primes are the chaotic trajectories that closed on themselves. However no hermitian operator is identified with the Riemann operator at hand, it is an analogy. Also the territory where the asymptotic expressions should work may locate much higher in the Riemann line $T\sim10^{100}$ far away of computing simulations. Billiard enclosures describing the zeros of Riemann do not have a the density of states as *log(T)*. From physics we know that this is very critical behaviour and billiards of dimension D just provide $T^{(D/2-1)}$ which is not logarithmic but a power law. Berry argued that there are some billiards [8] that may have the logarithmic behaviour, however this may be for D≥2. Another point is that the distribution of levels may have some long range oscillations over imposed to the Wigner function variance that may show in higher moments. In any case the work of Odlyzco and Berry set perspectives and excitation, at least for a physicist that may like to enter the game. It is nice idea: *Physical energy eigenvalues and eigenfunctions map the zeros of the zeta and the primes.* In particular my work is inspired in these ideas.

In this paper I would like to enter the game by searching for a one dimensional quantum mechanical Hamiltonian, defined by a hermitic operator, that will have a dispersion relation for the quantum number as that of formula (2). If I find such a hamiltonian and then link it to the Riemann function through an analytical operation, for example in a form of an integral equation, then I may have solved the unpublished Hilbert-Polya conjecture, that hopefully will sight some light into the Riemann hypothesis. The problem is not easy but is interesting and affordable. I have found such a Hamiltonian and the potential is the limiting harmonic potential that is defined by the *J(x)* function that is the generator of the prime number counting function *p(x)* that in turn is the integral over the zeta function [1]. The physical answer is that the potential of the one dimensional Hamiltonian is created by a force *J(x),* which spring of variable strength is the zeta function. The Hamiltonian is unique in one dimension. Therefore this hermitic operator may be assigned to the Hilbert–Polya operator and implies the conjecture to be true. The generalization to a two dimensional limiting harmonic potential of separated potentials in *x* and *y* (this will resemble the billiards) is straightforward. But does it imply the truth of the Riemann hypothesis? Not right away, but tell us the significance of going off the Riemann line.

The paper is distributed as follows: in section II I find the averaged and fluctuations of the Hamiltonian potential as well as the averaged values of the eigenvalues. In section III, I estimated the fluctuations of the eigenvalues that are due to the oscillating terms of the Riemann function. And finally I conclude the work in IV. Also the appendix A present the derivation of the potential.

## II. THE AVERAGED HAMILTONIAN POTENTIAL *(The approachissimo quantum harmonic oscillator)*

To treat the problem I will follow the strategy of finding the eigenvalues spectra of all the range of potential functions starting from logartithmic functions and going up to exponential functions. This way assures that if there is a Hamiltonian with a potential energy (from now on I switch from *T* to *E* because physicists we talk about energy) having the dispersion (2) I will chase it. There is no way that, if the potential exist, can escape if *V(x)* sweeps from *log(x)* to *exp(x)*. There is a point that I would like to mention from the start. For a physicist that observes the dispersion relation for energy as that of (2) is immediately clear that the averaged potential, without including the oscillations introduced by the roots or the primes, should be very near to the quantum harmonic oscillator. I fact the linear *E* (*T* in (2)) term is given by the harmonic oscillator and the *log(E)*, *log(T)*, should appear from a term weaker than $x^{2(1-e)}$ because this will give a power law for the energy dispersion relation. Therefore we already know that the potential has to be something like $x^2/(\log(x))^b$. We find that it is $\frac{x}{\log(x)} Li(x)$, (***b**=2 in first approximation*), where *Li(x)* is the logarithmic integral. **This result is detailed in appendix A**. This makes a lot of sense because one of the observations of the roots of the zeta function is that they show a repulsive behaviour. Of course there is nothing more repulsive that the harmonic potential that provided equiespaced levels; that is to say: its levels conforms a perfect crystal, perfect repulsion. The fact that we have a *log(E)* law implies that the repulsion is not perfect but the distance between levels (roots) is *1/log(E)*, almost constant, especially at high energies or short wavelengths. This is what switch the physical idea of energies with the mathematical function zeros. Therefore let us write the potential as

$$V_R(x) \approx \frac{x}{\log(x)} J(x) \qquad (3)$$

*J(x)* is the function introduced by Riemann [1] as,

$$J(x) = \frac{1}{2\boldsymbol{pi}} \int_{a-i\infty}^{a+i\infty} \log(\boldsymbol{z}(z)) \frac{x^z}{z} dz, \qquad a > 1 \qquad (4a)$$

or

$$V_R(x) = \frac{x}{2\boldsymbol{pi}\log(x)} \int_{a-i\infty}^{a+i\infty} \log(\boldsymbol{z}(z)) \frac{y^z}{z} dz, \qquad a > 1 \quad (4b)$$

and

$$J(x) = Li(x) - \sum Li(x^r) + Li(x^{\bar{r}}) + TNT \qquad (4c)$$

Where the TNT are two negligible terms [1], and ? are the roots of the Riemann function and I have taken the root and its complex conjugated for obtaining a real value of *J(x)* and of *$V_R(x)$*. Taking $\boldsymbol{r} = 1/2 + i\boldsymbol{a}$ is the Riemann hypothesis but the form of (4b) generalizes to non-Riemann hypothesis ($\boldsymbol{r} = \boldsymbol{b} + i\boldsymbol{a}$, with 0<ß<1) and *a* running to infinity. The formulae 4a and c are the main result of the celebrated paper by Riemann and from them the number of prime numbers $p(x)$ can be obtained.

Then we have $V_R(x) = V_{R0}(x) + V_{RP}(x)$. The terms under the sum in (4c) are a perturbation of $Li(x)$ and consist of the famous oscillating terms described by Riemann [1]. This summation represents a series conditionally converging and the way of summations is by taking pairs of complex conjugated roots from low to high latitudes. These are anomalous as Littlewood already proved it [9]. Then we have:

$$V_R(x) = V_{R0}(x) + V_{RP}(x) \qquad (5)$$

With $V_{R0}(x) = \dfrac{x}{\log(x)} Li(x)$ is the main or principal term of $J(x)$ that will be treated as the zero order of the Hamiltonian potential and $V_{RP}(x) = -\sum$ is the perturbative term of the potential to be treated in first order approximation and can be extended to higher order if desired. Let us define the potential $V_{R0}(x)$ more explicitly. This is:

$$V_{R0}(x) = \frac{x}{\log(x)} Li(x) \qquad \text{for } x \geq 2$$

$$V_{R0}(x) = 0 \qquad \text{for } 0 < x < 2 \qquad (6)$$

$$V_{R0}(x) = \infty \qquad \text{for } x \leq 0$$

The conditions for $x>0$ are perfectly satisfied by $J(x)$ by construction [1]. The infinite repulsive potential for $x<0$ implies zero penetration of the wave function in this region. This is a kind of half harmonic potential, plotted in Fig.1, and it can be extended to $x<0$ by taking the modulus of the variable. Then the prime number theorem can be expressed in terms of the zero order Riemann potential as:

$$\lim_{x \to \infty} p(x) \sim Li(x) \sim \frac{x}{\log(x)} \sim (V_{R0}(x))^{1/2} \qquad (7)$$

To proof that $V_{R0}(x)$ described by (6) has the averaged dispersion relation for the zero order $E_{N0}$ eigenvalues as in (2), we just take the quantum mechanical Schröedinger Hamiltonian

$$H_0 = -\frac{d^2}{dx^2} + V_{R0}(x) \qquad (8)$$

And solve for the eigenvalues $E_{N0}$ and eigenfunctions. And if one does this, it works. I do this by solving the differential equation, however if it is not easy to solve the differential equation we have an excellent technique called the WKB or semiclassical approximation [8,9] to the solutions of (9), especially this solution is very good at high energy or short wavelengths, the high $T$ the region where we are interested. And very interesting and accidentally, for the quantum harmonic oscillator, for the territory we are visiting, is an exact solution. This consists in calculating the integral of the momentum for the stationary trajectories and the eigenvalues satisfy

$$\int_0^{x_T} P(x)dx = \int_0^{x_T} (E - V(x))^{1/2} dx = 2p(N + 1/4) \qquad (9)$$

This is also the Sommerfeld quantization criterion and *N* is the quantum number of the solutions. The integral is done between the limits of the turning point of the potential given by $E = V(x_T)$. One of the turning points is $x_T=0$ and then imposes the wave function to be zero at *x=0* because the potential tends to infinity and then there is not quantum penetration of the eigenfunction for *x<0*. This together with the boundary conditions imposes the quantization values *2p (N+1/4)*. Notice that our approachissimo harmonic potential extends only for *x>0*, we could had symmetrised the potential for *x<0* also and then the quantization would have been *p (N+1/2)*. The *2p* appears because the increases in the kinetic energy due to the reduction in half of the quantum box width. The dispersion relation is:

$$N \sim \frac{E_{N0}}{2p} \log\left(\frac{E_{N0}}{2p} - 1\right) \qquad (10)$$

With the density of states,

$$w(E_{N0}) \approx dN/dE_{N0} \approx \log(E_{N0}) \qquad (11)$$

And the separation between levels is given by $w(E_{N0})^{-1} \sim 1/\log(E_{N0})$. The same behaviour as for the Riemann zeros (see equation (2)). Therefore eigenvalues and Riemann zeros can be identified. **(See appendix A).**

Therefore we have a 1-D quantum Hamiltonian $H_0$, obtained from the Riemann function, with eigenvalues that match the averaged number of zeros of the Riemann zeta function. We have transformed a number theory problem into a linear operator whose eigenvalues are known. Riemann took the counting of prime numbers into an integral equation and we are taking it into a physical quantum mechanics linear operator. So physical interactions are describing primality! Notice that the harmonic oscillator is the most common potential used in quantum mechanics. It describes vibration of atoms in solids, in molecules, in physisorbed species in surfaces, plasma frequencies, and all kind of modelling oscillators. However in these physical systems there are, in general, linear anharmonic terms and nevertheless in our potential describing the Riemann roots the anharmonic terms are much smaller, zero power law, logarithmic. So the most common hamiltonian of physics is related to the number theory of primality. This is a nice proof that physical reality is nothing but mathematics.

The wave functions in the WKB approximation that is very good for high energies, small wavelengths, and in particular these are exact for the harmonic potential:

$$\Psi_{N0}(x) = A\sin\left(\int_0^x \frac{(E-V(y))^{1/2} dy}{(E-V(x))^{1/4}}\right) \qquad 0 \le x \le x_T \qquad (13a)$$

$$\Psi_{N0}(x) = A\exp\left(-\int_{x_T}^x \frac{(E-V(y))^{1/2} dy}{(E-V(x))^{1/4}}\right) \qquad x > x_T \qquad (13b)$$

These two expressions correspond with the oscillatory behaviour with *N* nodes between the turning points (13a) and the exponential decay behaviour in the forbidden classical, quantum, region (13b). *A* is a normalization constant that is defined as:

$$\int_0^\infty |\Psi_{N0}(x)|^2 dx = 1 \qquad (14)$$

This integral is evaluated by expanding around the turning point $x_T \approx E_{N0}^{1/2} \log(E_{N0}^{1/2})$ that has an integrating root square singularity and using the stationary phase method, we obtain the value for *A*,

$$A^2 = \left\{ 2^{1/2} \log(x_T) + \left( \frac{2 x_T}{\log(x_T)^3} \right)^{-2/3} \right\}^{-1} \qquad (15)$$

In this expression the first term in *log(x)* comes from the main average part of the numerator in the integrand (14) and the smaller second term from the last oscillation of the potential around the turning point as well as from the exponential decay part of the wave function. Fig.1 shows a sketch of the wave function for a fixed energy. Therefore for large $E_{N0}$, $A^2$ goes as *1/log($x_T$)* and the evaluation of the integral is only between the turning points. This will have strong consequences in determining the eigenvalues $E_{N0}$ (containing the oscillatory terms of the potential). The cut-off that produces the exponential decay wavefunction in the forbidden classical region also cut-off the matrix elements of the perturbation that only have to be evaluated between the turning points [0 and $E_{N0} log(E_{N0})$]. A result that limits the evaluation of the Riemann oscillatory terms in the sum of (4c) to $x_T \approx E_{N_0}^{1/2} \log(E_{N_0}^{1/2})$; that is to say, to the eigenvalues in zero order, that may be taken also as the roots in zero order. The contribution of the exponential term (13b) to the first order perturbation matrix elements is just negligible as compared with the contribution of the oscillatory term (13a). We would like to notice that using WKB approximation for the harmonic oscillator provides and exact solution by accident. However we could also use for the zero order wave functions the corresponding to the solution quantum mechanical harmonic potentials. These are in our case, for the potential being infinity for *x<0*, the odd parity Hermite's orthogonal polynomial that vanish at *x=0*. So, the notes of the primes, if you would like to call it that way, are the Hermite polynomials of odd parity.

III. THE ROOTS FLUCTUATING POTENTIAL AND THE PERTURBATION THEORY.

We now proceed by calculating the matrix elements corresponding to first order perturbation theory with a perturbative potential given, as indicated above, by

$$V_{RP}(x) = \frac{x}{\log(x)} S(x) \qquad (16a)$$

$$S(x) = -\sum Li(x^r) + Li(x^{\bar{r}}) \qquad (16b)$$

We have the differential equation of eingenvalues:

$$[H_0(x) + V_{RP}(x)](\Psi_{N0}(x) + \Psi_{N1}(x)) = (E_{N0} + E_{N1})(\Psi_{N0}(x) + \Psi_{N1}(x)) \quad (17)$$

Taking terms to first order, $E_{N1}$ are given by the matrix elements,

$$E_{N1} = \langle \Psi_{N0} | V_{RP}(x) | \Psi_{N0}(x) \rangle = \int_0^{x_T} \Psi_{N0}(x) V_{RP}(x) \Psi_{N0}^*(x) dx \quad (18)$$

To perform this integral we expand the integrand around the turning point that has an integrable singularity, as we did to find the normalization coefficient $A$ (see (15)), and integrate between $x=0$ and $x=x_T$. The part of the integral from $x_T$ to infinity is negligible, because of the cut-off produced by the exponential decay wavefunction in the classical forbidden region, see discussion above. Of course this can also be estimated but gives terms much smaller by a power law of at least $x_T^{1/2}$. Then can be seen that the integral (18) over a function $V_{RP}(x)$ that contains powers of $x$ or $log(x)$, without singularities as, is our case, is just a linear application resulting in,

$$E_{N1} = \int_0^{x_T} \Psi_{N0}(x) V_{RP}(x) \Psi_{N0}^*(x) dx \to C \cdot V_{RP}(x_T, N0) \quad (19),$$

where $C$ is a constant to be determined. For the case at hand where $V_{RP}(x)=x^d g(lox(x))$, the constant $C=B(d+1,1/2)$, for the Riemann hypothesis $d=3/2$. Then the perturbed energy is just the application (19) to formula (16), but it is interesting to estimate also $V_{RP}(x)$ taking into account (19). The fact that the integral is cut-off from its infinite limit may have consequences from the point of view of the role played by the roots.

*i) Estimation of $V_{RP}(x)$*

In order to calculate the value of the perturbation potential, the first we have to define is the general behaviour of the roots. Here we are going to set a general behaviour and the roots will have the expression:

$$\mathbf{r}_i = a_i + i\mathbf{a}_i \quad (20)$$

Where we know that $0 < a_i < 1$ and $i$ runs from 1 to infinity. This produces the Riemann hypothesis when all $a_i = 1/2$. The sum (16) is over the values of $\mathbf{r}_i$ and its complex conjugated that guarantees us that the potential is a real valued function for $J(x)$. Then if we take the roots according with the graph in Fig.2, and considering the first term of the expansion of the $Li(x^r)$ integral, $S(x)$ reads,

$$S(x) = -\sum_i \left\{ \frac{x^{a_i}}{\log(x)} \cdot \frac{[2a_i \cos(\mathbf{a}_i \log(x)) + 2\mathbf{a}_i \sin(\mathbf{a}_i \log(x))]}{a_i^2 + \mathbf{a}_i^2} \right\} \quad (21a),$$

Taking into account Fig.2, we have

$$S(x) = -\frac{1}{\log(x)} \sum_{j=0}^{M_{j+1}} \sum_{i=M_j+1} x^{a_i(M_j)} \frac{[2a_i(M_j)\cos(\mathbf{a}_i(M_j)\log(x)) + 2\mathbf{a}_i(M_j)\sin(\mathbf{a}_i(M_j)\log(x))]}{a_i^2(M_j) + \mathbf{a}_i^2(M_j)} \quad (21b)$$

Notice that $a_i(M_j) << \mathbf{a}_i(M_j)$ for all $i$ and $j$ and therefore,

$$S(x) \approx -\frac{1}{\log(x)} \sum_{j=0}^{M_{j+1}} \sum_{i=M_j+1} x^{a_i(M_j)} \frac{[2a_i(M_j)\cos(\mathbf{a}_i(M_j)\log(x)) + 2\mathbf{a}_i(M_j)\sin(\mathbf{a}_i(M_j)\log(x))]}{\mathbf{a}_i^2(M_j)} \left(1 - \frac{a_i^2(M_j)}{\mathbf{a}_i^2(M_j)}\right)$$

(21c)

$M_0 = 0$.

Then the perturbation energies $E_{N1}$ are,

$$E_{N1} \approx B(d+1, 1/2).x_T / \log(x_T).S(x_T) \approx \mathbf{a}_N - E_{N0} \quad (22)$$

Equ.22 is an iterative equation for the $\mathbf{a}_N$'s and only enters the turning point. This equation acts as an attractor of the roots to the Riemann line due to the fact that the first large stretch of zeros is in this line. Assume that we are in a stretch of $a$ different of ½, then in this region the amplitud of the summands behaves as $x_T{}^a/(log(x_T)\mathbf{a}(a))$, but $\mathbf{a}(a) \gg x_T(a)^2 \gg E_N$, but the $x_T$ are determine the roots in the (0,1) strip. I have the impression that this keeps sending the roots to the ½ line. It is necessary to study equ.22 in a selfconsistent way although I do not have the techniques for this study.

The Riemman hypothesis is obtained for all $a_i = 1/2$. The value of the function $S(x_T)$ is a sum over all the roots that we do not know, but we can make a rough estimation. However we cannot obtain all its delicate structure without known the roots and adding them in (21b).

Its structure is similar to a Fourier transform although the interval $\mathbf{a}_i$ is not constant, however is almost constant and the values of $\mathbf{a}_i(M_j)$ are in the argument of a circular function riding a circle and varying fast. Then we can consider the sum to be transformed in an integral which discretization in the integrand is made by a non periodic interval. This is a crude approximation given the structure of the roots and will never recover the very short range fluctuation of the energies (roots), but will provide us with the asymptotic long range fluctuations. The $S(x)$ transforms into the integral,

$$S(x) \approx -\frac{1}{\log(x)} \sum_{j=0}^{M_{j+1}} \int_{i=M_j+1} \frac{2x^{a(M_j)}}{\mathbf{a}^2} \cdot \frac{[a(M_j)\cos(\mathbf{a}\log(x)) + \mathbf{a}\sin(\mathbf{a}\log(x))]}{} \cdot \left(1 - \frac{a^2(M_j)}{\mathbf{a}^2}\right) \log(\mathbf{a}) d\mathbf{a}$$

(21d)

In (21d) the *a*'s only depend now of the $M_j$ because in these intervals are constant and the **a** just run as the integrating variable. There is also the term *log(a)*, in the numerator of the integrand, that appears in passing from the discrete sum to the continuum, this is the density of states. The integrals of (21d) do not have a known primitive because of the *log(a)*, however we approximate it by considering *log(a)* at the limits of the integrals. This does produce a small error because the denominator behaves as $\alpha^2$. By taking this into account, and neglecting the term $\frac{a^2(M_j)}{\mathbf{a}^2}$, the integrals in (21d) have solutions in terms of *cos(**b**)* and *si(**b**)*. Then the function *S(x)* reads,

$$S(x) = -\frac{2}{\log(x)} \sum_{j=0} y^{a(M_j)} \left\{ \left[ si\left[\mathbf{a}(M_{j+1})\log(x)\right] - si\left[\mathbf{a}(M_j+1)\log(x)\right] \right] \cdot \left[1 + a(M_j)\log(x)\right] + \right.$$

$$\left. \frac{a(M_j)}{\mathbf{a}(M_j+1)} \left[ -\frac{\cos\left[\mathbf{a}(M_{j+1})\log(x)\right]}{\mathbf{a}(M_{j+1})} + \frac{\cos\left[\mathbf{a}(M_j+1)\log(x)\right]}{\mathbf{a}(M_j+1)} \right] \right\} \cdot \log(\mathbf{a}(M_j+1))$$

(21e)

And the Riemann hypothesis is obtained setting $a(M_j)=1/2$ for all $M_j$ and *j*.

$$S_R(x) = \frac{2y^{1/2}}{\log(x)} \left[ si(\mathbf{a}_1 \log(x)) + \frac{1}{2}\log(x) si(\mathbf{a}_1 \log(x)) - \frac{1}{2\mathbf{a}_1}(\cos(\mathbf{a}_1 \log(x))) \right] \log(\mathbf{a}_1)$$

(21f)

In the above expressions we notice that because $\mathbf{a}_1 \sim 14.2$ and the minimum value of *log(x)* is *log(2)* we have that the minimum argument of the *si* function is around $4\mathbf{p}$ therefore we do not produce a large error by taking the asymptotic expression for the *si* function. In this way we have $S_R(x)$, according to (16a), for the Riemann hypothesis $V_{RRP}(x)$,

$$E_{N1} = -\frac{3\mathbf{p} \cdot x_T^{3/2}}{4 \cdot \log(x_T)^2} \left[ \cos(\mathbf{a}_1 \log(x_T)) + O(\frac{1}{\log(x_T)} \cos(\mathbf{a}_1 \log(x_T))) \right] \frac{\log(\mathbf{a}_1)}{\mathbf{a}_1} \qquad (22)$$

Fig.3 shows a plot of the energies $E_{N0}$ and $E_T = E_{N0} + E_{N1}$. It can be observed that the last energy presents oscillations of large period. These appear as consequence of passing the sums to integrals, however we think that the long period oscillations, that increase with *N*, may not be a good approximation for low values of *N* because the roots are non periodic. When the "randomness" in the roots is introduced these oscillations should tend to disappear because there many interdistances between roots mixed and in addition we have used asymptotic conditions. Nevertheless for very large *N*, large energies (roots in the high latitudes, much higher that those values of the Fig.2), these oscillations should remain and should be observable and are independent of the Riemann hypothesis, given the extgructure of *S(x) and equ, 22*. They have an influence in the roots and also in the primes and they may be related to the problem already solved and discussed by Littlewood [9] that the conjecture $p(x)<Li(x)$ is false. In fact for asymptotic $N$ there is an infinite number of oscillations where the above inequality fails.

The oscillations increase their amplitude with $N$, as $\dfrac{3p \cdot x_T^{3/2}}{4 \cdot \log(x_T)^2} \dfrac{\log(a_1)}{a_1}$, however their relative value with respect to $E_{N0}$ tends to zero at least as $x^{-1/2}$.

As discussed in the Appendix A due to precision to estimated the integrals of eingenvalues we cannot distinguish clearly of the $V_R(x) \approx x/\log(x)Li(x)$ or $\dfrac{2}{\log(x)}\int_2^x J(x)dx$. Within our estimations, both give practically the same dispersion relation (10). By proceeding as above and using the integral expression for the potential, we obtain,

$$E_{N1} = -\dfrac{2^{3/4}3p}{4\left(\dfrac{9}{4}+a_1^2\right)\log(x_T)}\left[F_p(x_T)-F_p(2)\right]\dfrac{\log(a_1)}{a_1} \qquad (23a)$$

With

$$F_p(x_T) = -\dfrac{x_T^{3/2}}{\log(x_T)}\left[\dfrac{3}{2}\cos(a_1\log(x_T))+a_1\sin(a_1\log x_T)\right]\left[1+O\left(\dfrac{1}{\log(x_T)}\right)\right]$$
(23b).

Formula (23) shows the same type of long-range oscillations but a factor of 100 approximately smaller. But from the point of view of the asymptotic behaviours this is the same result.

IV CONCLUSSIONS

1. As we have seen the hermitic physical hamiltonian that maps the zero of the zeta function zero exists but this does not implies that the zeros lie in the line $x=1/2$. Any distribution of complex conjugated zeros in the strip delimited by $0<x<1$ provides the hermitic operator. However the introduction of zeros off the critical Riemann line affects the primality function $p(x_T)^2 \approx V_R(x_T)$ in a way that depends very much on the way that the number of zeros stretched out of the line, but it looks as if the ½ line is an attractor of the zero due to the iterative equation 22. All the relations between numbers and the physics take place at the turning point of the potential. This is like a track where the primes move. This potential is an *approachissimo* harmonic potential that is the most common potential in physics. The harmonicity is so pure that the anharmonicity departs from it logarithmically, zero power law. This implies, for example, that a solid made with such a potential interaction will have a completely negligible thermal expansion no matter how much the temperature increases. The solid will not melt.
2. The problem has been studied in 1-D but the generalization to 2-D is just straightforward by adding to the Hamiltonian the separate Hamiltonian in the *y* variable. This will be a billiard as those described by Berry [6,7] and will have chaotic trajectories. The potential is a 2-D paraboloid function corrected by the logarithmic function as described above.
3. There are very long range oscillations in the eigenvalues (roots) distribution that makes the **p**(x) oscillate above and below *Li(x)*. A result already proven by Littlelwood and that here appears from a simple estimation of the eigenvalues. These oscillations are independent of the Riemann hypothesis.
4. Finally, I would like to say paraphrasing Berry: *Yes, there is harmony and music in the primes. And it looks to us as if the notes are the Hermite Orthogonal Polynomials with slightly off undertones. The primes are harmony described by an approachissimo harmonic physics world. The spring of this approachissimo is the zeta Riemann function.*

**APENDIX A**

*Determination of the Zero Order Potential $V_{R0}(x)$.*

In this appendix we proceed by determining the expression for $V_{R0}(x)$ that satisfy the dispersion relation (10),

$$\frac{E_{N0}}{2p}\log\left(\frac{E_{N0}}{2p}-1\right) \sim N \qquad (A1)$$

As we mention in the previous text, this is done by testing all kind of potential functions $G(x)$ starting in the $log(x)$ and moving up to $e^x$. The potential will have the general form:

$V_{R0}(x) = G(x)$          for    $x \geq 0$      (A2a)

$V_{R0}(x) = \infty$          for    $x < 0$      (A2b)

Fig.1 shows a sketch of the potential with the wave function shape, the turning points $x_T$ and the wave function for a given energy. The eigenvalues are obtained by using the WKB method and when possible integrating the corresponding differential equation for the Hamiltonian $H_0$ (8),

$$H_0 = -\frac{d^2}{dx^2} + V_{R0}(x) \qquad (A3)$$

The WKB consists in integrating the momentum trajectories with the boundary conditions,

$$\int_0^{x_T} P(x)dx = \int_0^{x_T} (E-V(x))^{1/2} dx = 2p(N+1/4) \qquad (A4)$$

    i)      *The logarithmic potential $G(x) = log(x)$.*

In this case the solution of the equation for (A3) is not known, at least to us, and the integral for (A4) is a Beta function. The dispersion relation is given by,

$$e^E \gg N \qquad (A5)$$

    ii)      *The linear potential $G(x) = x$*

The solutions are the Airy functions [12] and the WKB correspond with the same dispersion relation for high *N*. This potential appears, for example, in the localization of electrons in surfaces with an applied electric field [13].

$$E^{3/2} \gg N \qquad (A6)$$

*iii) The cuadratic harmonic oscillator potential, $G(x) = x^2$*

The solutions to the Schroedinger´s equation are the Hermite orthogonal polynomials of odd parity because the wave function vanishes at *x=0*. Accidentally, in this case the WKB and the exact solutions are the same. This is the most common used potential in physics, as mentioned above.

$$E \gg N \qquad (A7)$$

iv)     *The exponential potential, $G(x) = e^x$*

The solutions to the Schroedinger´s equation are combinations of Bessel functions[14]. The WKB solution is the same for large *N*. This potential appears as an approximation to the potential that localize adsorbed and chemisorbed atoms at surfaces.

$$log(E) \gg N \qquad (A8)$$

We have also check hydrogenic type of potentials that localize images states at surfaces [15], but none of these can satisfy the dispersion relation (A1). It is clear then from the above results that the needed potential is very near to the harmonic oscillator potential (iii).

v)     *Approaching the harmonic oscillator $G(x)=x^{2-e}$, (**e** small)*

The WKB integral gives a beta function,

$$E^{1+4e} \gg N \qquad (A9)$$

The dispersion relation is a power law, stronger than the *log(E)* of (A1). Therefore the potential has to approach more to the harmonic potential. In fact the *log* term implies approaching the harmonic potential in a zero power fashion.

vi)     **Aproachissimo** *harmonic oscillator $G(x)=x^2/log(x)^b$*

   a)     **b**=1;

The WKB integral does not have a known primitive and we have to estimate it by other methods. We do this by expanding the integrand (A4) around the turning point, as follows:

$$x_T = aE^{1/2}, \text{ then } 1 = \frac{a^2}{\log(E^{1/2}) + \log(a)}.$$ By expanding the integral around $x_T$ we have,

$$E \log(E)^{1/2} \approx N \qquad (A10a)$$

We see that **b=1** is too approachisimo to the harmonic.

b) **b=2;** $G(x) = x^2/\log(x)^2$

Proceeding as in *(a)* we have,

$$E[\log(E) + \log(\log(E))] \approx N \qquad (A10a)$$

This is a little too far from the harmonic to obtain our desired dispersion relation. We need a series expansion that brings us near to the dispersion relation (A1).

c) **Connecting with the Riemann J(x) function**

As discussed in *section II* the main principal part of $J(x)$ is the logarithm integral $Li(x)$,

$$Li(x) = \int \frac{dt}{\log(t)} = \frac{x}{\log(x)} + \frac{x}{(\log(x))^2} + 2\int \frac{dx}{(\log(x))^3} - Li(2) \qquad (A11),$$

Now, if we define

$$G(x) = \frac{x}{\log(x)} Li(x) = x^2 \left( \frac{1}{(\log(x))^2} + \frac{1}{(\log(x))^3} \right) + \text{lower order terms} \qquad (A12)$$

We have an expansion of the case **b=2**. To estimate the WKB integral, proceed as above, *(iv.a)*, and expand the integral a round the turning point, then integrate. This results in the dispersion relation

$$E \cdot \left[ \log(E) - 1 + O\left[ \log\left( \frac{\log(E)}{\log(E)} \right) \right] \right] \approx N \qquad (A13)$$

The next iteration in the series probably cancels the *log(log(E))*, however we are also having errors in the estimation and is not worth to continue.

Finally we shall mention that if we take for $G(x)=2/log(x) \int_2^x J(y)dy$, the result within the error would have give the same one as in (A14). However the two functions are approximately the same and in first order they coincide.
**Therefore we take $G(x)=x/log(x) \approx Li(x)$ for defining $V_{R0}(x)$ and by extension $V_R(x) \approx x/log(x) \approx J(x)$.**

## FIGURE CAPTIONS

Figure 1.   Plot of the potential and the wave function versus $x$ for a given energy. The turning points are also indicated.

Figure 2.   Sketch of the roots distribution in the strip (0,1). The imaginary part of the roots are distributed in the thick lines that have mirror symmetry with the line $a=0$. These are complex conjugated and their sum yields real values for J(x) and the potential provides a hermitic hamiltonian.

Figure 3.  Plot of the energies $E_{N0}$ and $E_{N0}+E_{N1}$ versus $N$. Notice that the $E_{N1}$ values have been multiplied by 30 to stress the amplitude of the long-range oscillations. The range in the plot does not imply that the oscillations may appear for these small values of the energies but is just as an example. We expect that when adding the real roots these oscillations may appear at much higher energies.

*REFERENCES*

1. H. M. Edwards, *Riemann´s Zeta Function*, Academia Press (New York, 1974).
2. M. du Sautoy, *The Music of the Primes*, Harper Perennial (Printed by Harper Collins Publishers), London, 2004.
3. J. Derbyshire, *Prime Obsession*, Joseph Henry Press (Washington D.C. 2001).
4. For the never published Hilbert-Polya conjecture, see the web of A.M. Odlyzko.
5. A. M. Odlyzko, Math. Comp. **48**, 273(1987).
6. M.V. Berry, in *Quantum Chaos and Statistical Nuclear Physics (Springer Lecture Notes in Physics, N.263)*, eds. T. H. Seligman and H. Nisioka (Berlin 1986).
7. M. V. Berry, Nonlinearity **1**, 399 (1988).
8. B. Simon, Ann. Phys.**146**, 2009 (1983).
9. J. E. Littlewood, C.R. Acad. Sci. Paris, **158**, 1869 (1914).
10. A. Messiah, *Mecánica Cuántica*, (Tecnos , Madrid 1965), Chap.VI.
11. R. E. Langer, Phys. Rev.51, 669(1937).
12. W. Magnus, F. Oberhettinger and R. P. Soni, *Formulas and Theorems for the Special Functions of Mathematical Physics,* Springer-Verlag, (Berlin-Heidelberg 1966).
13. G. Binnig, K. H. Frank, H. Fuchs and N. García, Phys. Rev. Lett. **45**, 991 (1985).
14. N. García and J. Ibáñez, J. Chem. Phys. **64**, 4803 (1976).
15. N. García and J. Solana, Surface Sci. **36**, 262 (1973).

**FIGURES**

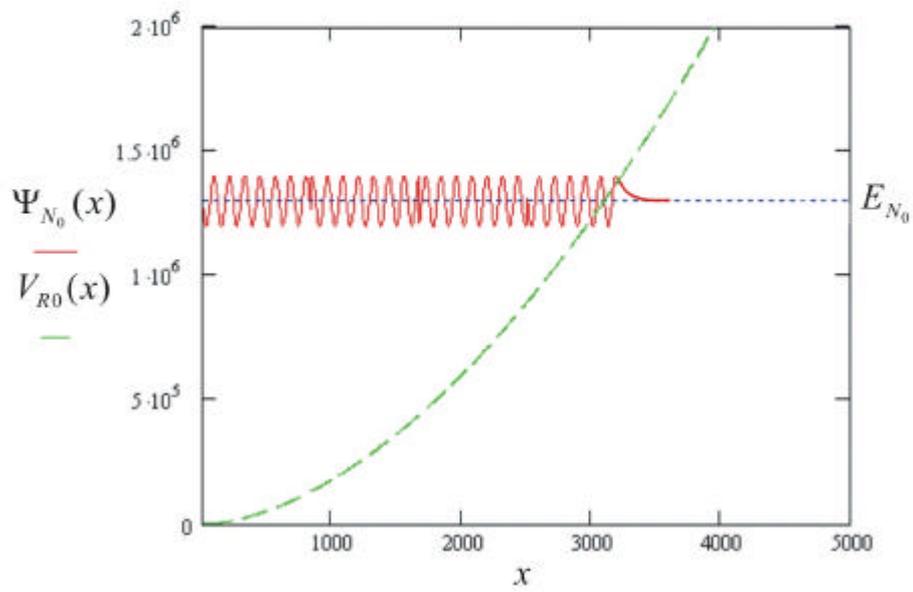

fig. 1

fig. 2

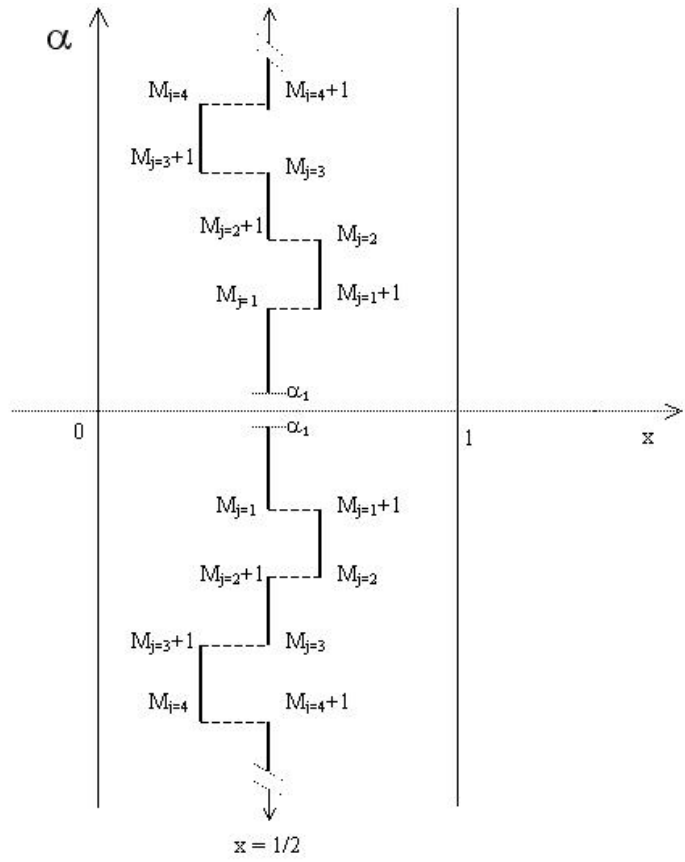

Figura 2

fig. 3

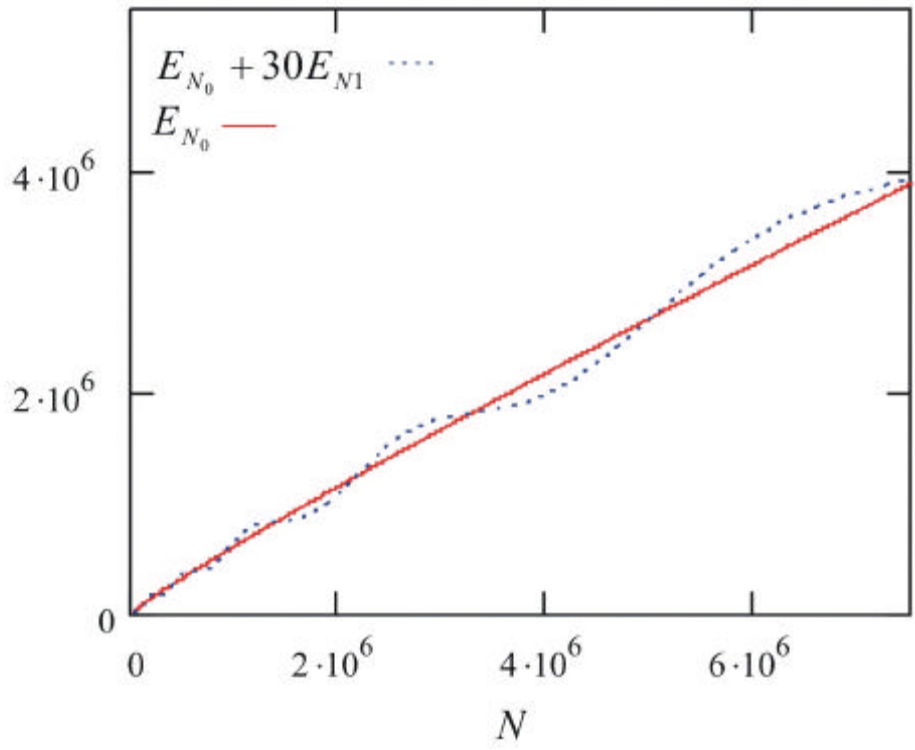